\documentclass{PoS}

\usepackage{latexsym}

\usepackage{amssymb,amsfonts,amsmath}
\usepackage{graphicx} 
\usepackage{indentfirst}

 \usepackage{bbm}

\newcommand {\cB}{{\cal B}}
\newcommand {\cC}{{\cal C}}
\newcommand {\cD}{{\cal D}}

\newcommand {\cK}{{\cal K}}

\newcommand {\cM}{{\cal M}}
\newcommand {\cN}{{\cal N}}

\newcommand {\cR}{{\cal R}}
\newcommand {\cS}{{\cal S}}
\newcommand {\cT}{{\cal T}}

\newcommand {\cW}{{\cal W}}

\newcommand{\bD}{{\bf D}}

\def\a{\alpha}

\def\b{\beta}

\def\d{\delta}
\def\e{\epsilon}
\def\f{\phi}
\def\g{\gamma}

\def\k{\kappa}
\def\l{\lambda}
\def\m{\mu}

\def\o{\omega}

\def\q{\theta}
\def\r{\rho}
\def\s{\sigma}
\def\t{\tau}

\def\x{\xi}
\def\z{\zeta}

\def\F{\Phi}
\def\J{\Psi}
\def\L{\Lambda}
\def\O{\Omega}

\def\S{\Sigma}
\def\U{\Upsilon}
\def\X{\Xi}

\def\rd{{\rm d}}
\def\ri{{\rm i}}
\def\re{{\rm e}}

\newcommand{\ad}{{\dot{\alpha}}}                           
\newcommand{\bd}{{\dot{\beta}}}                            
\newcommand{\ve}{\varepsilon}                            
\newcommand{\cDB}{{\bar\cD}}                            

\newcommand{\ab}{{\a\b}}

\newcommand{\pa}{\partial}                           
\newcommand{\hf}{\frac12}

\newcommand{\vf}{\varphi}

\newcommand{\be}{\begin{equation}}
\newcommand{\ee}{\end{equation}}
\newcommand{\bea}{\begin{eqnarray}}
\newcommand{\eea}{\end{eqnarray}}
\newcommand{\non}{\nonumber}
\def\double #1{#1{\hbox{\kern-2pt $#1$}}}

\newcommand{\hm}{{\hat{m}}}
\newcommand{\hn}{{\hat{n}}}

\newcommand{\ha}{{\hat{a}}}
\newcommand{\hb}{{\hat{b}}}

\newcommand{\hal}{{\hat{\a}}}

\newcommand{\bsubeq}{\begin{subequations}}
\newcommand{\esubeq}{\end{subequations}}

\title{Supersymmetric Spacetimes from Curved Superspace}

\ShortTitle{Symmetries of Curved Superspace}

\author{\speaker{Sergei M. Kuzenko}
\\
School of Physics M013, The University of Western Australia\\
35 Stirling Highway, Crawley W.A. 6009, Australia\\
        E-mail: \email{sergei.kuzenko@uwa.edu.au}}

\abstract{We review the superspace technique to determine supersymmetric 
spacetimes in the framework of off-shell formulations for supergravity 
in diverse dimensions using the case of 3D $\cN=2$ supergravity theories  
as an illustrative example. This geometric formalism has several advantages 
over other approaches advocated in the last four years.
Firstly, the infinitesimal isometry transformations of a given curved superspace 
form,  by construction, a  finite-dimensional Lie superalgebra, with its odd part corresponding to the rigid  supersymmetry transformations. 
Secondly, the generalised Killing spinor equation, 
which must be obeyed by the supersymmetry parameters, is a consequence of 
the more fundamental superfield Killing equation. 
Thirdly, general rigid supersymmetric theories on a curved spacetime 
are readily constructed in superspace by making use of the known 
off-shell supergravity-matter couplings and restricting them to the background chosen. 
It is the superspace techniques which make it possible to generate arbitrary 
off-shell supergravity-matter couplings. 
Fourthly, all maximally supersymmetric Lorentzian spaces correspond to those 
off-shell supergravity backgrounds 
for which the Grassmann-odd components 
of the superspace torsion and curvature tensors vanish, while the Grassmann-even components of these tensors are annihilated by the spinor derivatives. 
}

\FullConference{Proceedings of the Corfu Summer Institute 2014 "School and Workshops on Elementary Particle Physics and Gravity",\\
		3-21 September 2014\\
		Corfu, Greece }

\begin{document}

\section{Introduction}

Supersymmetric solutions of supergravity theories  had already
attracted much interest in the early 1980s in the context of Kaluza-Klein
supergravity, see \cite{DNP} for a review. At that period, 
the notion of Killing spinors\footnote{F. Englert, one of the authors
of  \cite{ERS}, was awarded the 2013 Nobel Prize in Physics (shared with P. Higgs).}
\cite{DP,ERS} (see also \cite{vanNW}), which  
is crucial to the program of Kaluza-Klein supergravity, was introduced. 
The existence of such spinors guarantees some
unbroken supersymmetry upon compactification. 
Since then, the Killing spinors on pseudo-Riemannian manifolds and
their properties have been studied by mathematicians, 
see \cite{BFGK,Friedrich} and references therein.  

An additional impetus to study supersymmetric solutions 
of supergravity theories 
comes from superstring theory to which supergravity is 
a low-energy approximation. 
Due to certain non-renormalisation and stability properties they possess,
such solutions are of special importance in the string-theoretic framework. 
A detailed discussion of the huge number of the supersymmetric solutions 
of supergravity constructed in diverse dimensions
is beyond the scope of this conference paper.
As an example of such constructions, it is pertinent to mention two papers
\cite{GG1} in which all supersymmetric solutions in minimal Poincar\'e and anti-de Sitter supergravity theories in five dimensions were constructed. 

In off-shell supergravity, the superspace formalism to determine (super)symmetric back\-gro\-unds was elaborated twenty years ago \cite{BK} in the framework 
of the old minimal formulation \cite{WZ,old}
 for $\cN=1$ supergravity in four dimensions (4D). 
 The approach developed in \cite{BK} is  universal, for  
 it may  be generalised to derive supersymmetric backgrounds 
 associated with any supergravity theory formulated in superspace.
In particular, it has already been used to construct rigid supersymmetric field theories
in 5D $\cN=1$ \cite{KT-M07}, 4D $\cN=2$ \cite{KT-M08,BKads,BKLT-M}
and 3D $(p,q)$   \cite{KT-M11,KLT-M12,BKT-M} anti-de Sitter superspaces. 

Recently,  much progress has been made in deriving new exact results 
for observables (partition functions, Wilson loops) 
in rigid supersymmetric gauge theories on compact manifolds such as round spheres 
using localisation techniques \cite{Pestun,KWY,Jafferis,HHL}. 
In order to apply these techniques, 
two technical prerequisites are required. Firstly, a curved space $\cM$  
has to admit some unbroken rigid supersymmetry. 
Secondly, the rigid supersymmetric theory on $\cM$ should be off-shell. 
These conditions are met by those supersymmetric backgrounds 
that correspond to off-shell supergravity theories. 
This is why a number of publications have appeared which 
are devoted to the construction 
of supersymmetric backgrounds associated with  off-shell supergravity theories 
in diverse dimensions, see \cite{FS,Jia:2011hw,Samtleben:2012gy,Klare:2012gn,DFS,
KMTZ,Liu:2012bi,Dumitrescu:2012at,Kehagias:2012fh,Festuccia3D,HTZ,deMH1,
DKSS2}
and references therein.   
Inspired by 
the work of Festuccia and Seiberg
\cite{FS}, these authors used component field considerations. 
In the case of 4D $\cN=1$ supergravity, it was shown 
\cite{K13} how to derive the key component results of, e.g.,  \cite{FS,KMTZ} from the 
more general superspace construction of \cite{BK}.
Recently, the formalism of \cite{BK,K13} was extended to construct supersymmetric backgrounds \cite{KLRST-M} 
associated with all known off-shell formulations for 
3D $\cN=2$ supergravity \cite{KT-M11,KLT-M11}.
The results obtained are in agreement with the component considerations of 
\cite{Festuccia3D,HTZ,DKSS2}.
The same formalism has also been used in \cite{KNT-M} 
to derive supersymmetric backgrounds in off-shell formulations 
for 5D $\cN=1$ supergravity. 

In this paper, we give a pedagogical review of the formalism of \cite{BK,K13}. 
As an application of the formalism, we briefly describe the results of
\cite{KLRST-M} devoted to the construction of supersymmetric 
backgrounds in  all known off-shell formulations for 
3D $\cN=2$ supergravity \cite{KT-M11,KLT-M11}.

\section{(Conformal) isometries of curved space}

Before discussing supersymmetric backgrounds in off-shell supergravity,  
it is instructive to recall how 
the (conformal) isometries of a curved spacetime 
are defined within the Weyl invariant formulation for gravity 
\cite{Deser,Zumino,Dirac}.  Our presentation follows \cite{KNT-M}.
We start by recalling three known approaches to the description of gravity 
on a $d$-dimensional manifold $\cM^d$: (i) the metric formulation; 
(ii) the vielbein formulation; and (iii) the Weyl invariant formulation.
 
In the standard metric approach, the gauge field is a symmetric metric 
tensor $g_{mn} (x) $ 
such that $g:= \det (g_{mn}) \neq 0$. 
The infinitesimal gauge transformation of $g_{mn} $  is 
\bea
 \d g_{mn} =  \nabla_m \x_n + \nabla_n \x_m~,
 \eea
 with  the gauge parameter $\x = \x^m (x) \pa_m$ 
 being a vector field generating 
a one-parameter family of diffeomorphisms.
 
In the vielbein formulation,
the gauge field is  a vielbein $e^a := \rd x^m e_m{}^a (x)$ that constitutes an orthonormal  
basis in the cotangent space 
$T^*_x \cM^d$,  for any spacetime point $x$, 
  $e:=\det (e_m{}^a ) \neq 0$.
The metric becomes a composite field defined by $g_{mn} = e_m{}^a e_n{}^b \eta_{ab} $, 
with $\eta_{ab}$ the Minkowski metric. 
The gauge group is now larger than in the metric approach. 
It includes both general coordinate and local Lorentz transformations, 
\bea
\d \nabla_a = [ \x^b \nabla_b +\hf K^{bc}M_{bc} , \nabla_a]~,
\label{2.2}
\eea
with the gauge parameters $\x^a (x) = \x^m (x) e_m{}^a (x) $ and 
$K^{ab} (x) = - K^{ba}(x)$ 
being completely arbitrary.
The gauge transformation \eqref{2.2} makes use of 
the torsion-free covariant derivatives, 
\bea
\nabla_a = e_a +\o_a = e_a{}^m  \pa_m +\hf \o_a{}^{bc}  M_{bc} ~, \qquad 
[\nabla_a , \nabla_b ] = \hf R_{ab}{}^{cd} M_{cd} ~.
\label{2.33}
\eea
Here $M_{bc} = -M_{cb}$ denotes the Lorentz generators, 
$ e_a{}^m (x) $ the inverse vielbein, $e_a{}^m e_m{}^b = \d_a{}^b$,  
and $\o_a{}^{bc} (x)$ the torsion-free Lorentz connection. Finally, 
$R_{ab}{}^{cd} (x)$ is the Riemann curvature tensor; 
its descendants are the Ricci tensor 
$R_{ab} := \eta^{cd} R_{c a db} = R_{ba}$ and the scalar curvature
$R=\eta^{ab}R_{ab}$. 

As is well-known, 
the torsion-free constraint
\bea
T_{ab}{}^c=0 \quad \Longleftrightarrow \quad 
[\nabla_a , \nabla_b ] \equiv T_{ab}{}^c \nabla_c +\hf R_{ab}{}^{cd} M_{cd} 
=\hf R_{ab}{}^{cd} M_{cd}
\label{torsion-free}
\eea
is invariant under  Weyl (local scale) transformations of the form
\bea
\nabla_a \to \nabla'_a =  \re^{\s} \Big( \nabla_a +(\nabla^b\s) M_{ba}\Big)~,
\label{A.5}
\eea
with the parameter $\s(x)$ being completely arbitrary.  
Such a transformation is induced by that of the gravitational field 
\bea
 e_a{}^m \to \re^\s e_a{}^m 
\quad \Longrightarrow \quad g_{mn} \to \re^{-2\s}g_{mn} ~.
\eea

In general, Weyl invariant matter theories are curved-space extensions 
of ordinary conformally invariant theories. 
As an example, consider the model for a scalar field $\vf$ with action
\bea
S=-\hf \int \rd^d x \,e \,\Big\{ \nabla^a \vf \nabla_a \vf +\frac{1}{4} \frac{d-2}{d-1} 
R \vf^2  +{\l} \vf^{2d/(d-2)} 
\Big\}~, \qquad d\neq 2~,
\label{A.7}
\eea
where $\l$ is a coupling constant. 
The action is Weyl invariant\footnote{The
Weyl transformation 
of the scalar curvature 
is $ R\to \re^{2\s} \Big\{ R + 2(d-1) \nabla^a\nabla_a \s 
- (d-2)(d-1) ( \nabla^a \s ) \nabla_a \s\Big\}$.}
provided 
$\vf$ transforms as 
\bea 
\vf \to \vf' = \re^{\hf (d-2)\s} \vf~. 
\label{A.8} 
\eea
The flat-space limit of \eqref{A.7} is a conformal field theory. 

Most field theories in curved space do not possess Weyl invariance.
In particular, the  pure gravity action with  a cosmological term
\bea
S_{\rm GR}= \frac{1}{2\k^2} \int \rd^d x \,e \,  R   -\frac{\L}{\k^2} \int \rd^dx \,e 
\label{A.9}
\eea
is not invariant under the Weyl transformations \eqref{A.5}.
However, any field theory in curved space can be made Weyl invariant by coupling it 
to a conformal compensator.

In the Weyl invariant formulation for gravity in $d\neq 2$ dimensions,
the gravitational field is described in terms of two gauge fields. 
One of them is the vielbein $e_m{}^a  (x)$ and the other is 
a conformal compensator $\vf (x)$
with the Weyl transformation law \eqref{A.8}. 
As compared with the matter model \eqref{A.7}, 
the compensator is constrained to be nowhere 
vanishing, $\vf \neq 0$.  In this approach, the gravity gauge group is defined to consist 
of the general coordinate,
local Lorentz and Weyl transformations
\begin{subequations}\label{A.10}
\bea
\d \nabla_a &=&  [ \x^b \nabla_b +\hf K^{bc}M_{bc} , \nabla_a]
+  \s \nabla_a +(\nabla^b\s) M_{ba} \equiv ( \d_\cK + \d_\s )\nabla_a~,  \label{A.10a} \\
\d \vf &=&  \x^b \nabla_b \vf + \hf (d-2)\s \vf \equiv ( \d_\cK + \d_\s ) \vf~ , 
\eea
\end{subequations}
where we have denoted $\cK:= \x^b \nabla_b +\hf K^{bc}M_{bc} $.
Any dynamical system is required to be invariant under these 
transformations.
 In particular, the Weyl invariant gravity action is 
 \bea
S_{\rm GR}=\hf \int \rd^d x \,e \,\Big\{ \nabla^a \vf \nabla_a \vf +\frac{1}{4} \frac{d-2}{d-1} 
R \vf^2  +{\l} \vf^{2d/(d-2)} 
\Big\}~.
\label{A.11}
\eea
Applying a finite Weyl transformation allows us to choose the gauge condition
\bea 
\vf = \frac{1}{2\k}  \sqrt{\frac{d-1}{d-2}}~,
\eea
in which the action \eqref{A.11} turns into \eqref{A.9}.  

Every gravity-matter system can be made Weyl invariant by 
replacing  $e_a{}^m \to \vf^{-2/(d-2)}e_a{}^m$ in the action. 
If the action of a Weyl invariant theory does not depend on $\vf$, 
it describes conformal gravity coupled to matter. 
It is natural to use the notation $(\cM^d, \nabla) $ in the case of conformal gravity
and $(\cM^d, \nabla, \vf) $ for ordinary gravity. In both cases, the gravity gauge freedom 
is given by \eqref{A.10}, but $\vf$ is not present in conformal gravity. 
One may understand conformal gravity as possessing an additional gauge freedom, 
$\nabla_a \to \nabla_a$ and $\vf \to \re^\r \vf$, 
with the gauge parameter $\r(x)$ being arbitrary.  

Let us fix a background spacetime. 
A vector field $\x = \x^m \pa_m = \x^a e_a$ on  $(\cM^d, \nabla) $
is  said to  be conformal Killing if there exist local Lorentz $K^{bc}[\x]$ and 
Weyl $\s [\x] $ parameters such that 
\bea
\big( \d_{\cK[\x]} + \d_{\s[\x]} \big)\nabla_a= 0~.
\label{cKvf}
\eea
A short calculation gives
\bea
K^{bc} [\x] &=& \hf \big( \nabla^b \x^c - \nabla^c \x^b \big)~, \qquad
\s[\x] = \frac{1}{d} \nabla_b \x^b  ~,
\eea
as well as the conformal Killing equation
\bea
 \nabla^a \x^b + \nabla^b \x^a = 2 \eta^{ab} \s[\x] ~.
 \eea
The set of all conformal Killing vector fields on
$(\cM^d, \nabla )$
forms a finite-dimensional Lie algebra.\footnote{Introducing
$\U := \big\{ \x^b, K^{bc}[\x] , \s[\x], \nabla_b \s[\x] \big\}$, one observes 
that $\nabla_a \U \subset {\rm span}(\U)$.}
It is the conformal algebra of the spacetime, and 
its dimension cannot exceed that of ${\frak so}(d,2)$, 
the conformal algebra of Minkowski space. 
The notion of a conformal Killing vector field does not make use of $\vf$, and therefore
two spacetimes $(\cM^d, \nabla, \vf )$ and $(\cM^d, \nabla, \re^\r \vf )$ 
have the same conformal Killing vector fields, for an arbitrary scalar field $\r(x)$. 
 
Two spacetimes 
$(\cM^d, \nabla, \vf )$ and $(\cM^d, \widetilde{\nabla},  \widetilde{\vf} )$
are said to be conformally related 
if their gauge fields are obtained from each other by applying a finite Weyl transformation, 
\bea
\widetilde{\nabla}_a &=&  \re^{\r} \Big( \nabla_a +(\nabla^b\r) M_{ba}\Big) ~,\qquad
\widetilde{\vf} = \re^{\hf (d-2)\r} \vf ~,
\label{A.16}
\eea
for some $\r$. These spacetimes 
have the same conformal Killing vector fields, 
\bea 
\x = \x^a e_a = \tilde{\x}^a \tilde{e}_a~.
\eea 
The parameters $K^{cd}  [\tilde{\x} ] $ and $\s [ \tilde{\x} ] $ are related to 
$K^{cd}  [{\x} ] $ and $\s [ {\x} ] $ as follows:
\bea
\cK [\tilde{\x} ] &:=& \tilde{\x}^b \widetilde{\nabla}_b + \hf K^{cd}[\tilde{\x} ] M_{cd}
=\cK[\x]~, \\
\s [ \tilde{\x} ] &=& \s [ \x ] - \x \r ~.
\eea
These relations are such that 
$( \d_{\cK[\tilde\x]} + \d_{\s[\tilde \x]} )\widetilde{\nabla}_a= 0$.

A conformal Killing vector field $\x = \x^a e_a$ on $(\cM^d, \nabla, \vf )$
is  called Killing 
if the transformation $\d_{\cK [\x]}  + \d_{\s [\x]}$
 does not change the compensator,  
\bea
\x \vf +\hf (d-2) \s[\x] \vf &=&0 ~.\label{A.19}
\eea
The set of all  Killing vector fields of the given spacetime $(\cM^d, \nabla, \vf )$
is a finite-dimensional Lie algebra. By construction, it is a subalgebra of
the conformal algebra of $(\cM^d, \nabla )$. 
The Killing equations \eqref{cKvf} and 
\eqref{A.19} are Weyl invariant in the following sense.
Given a conformally related spacetime $(\cM^d, \widetilde{\nabla}_a,  \widetilde{\vf})$
defined by eq.~\eqref{A.16},
the  Killing equations \eqref{cKvf} and  \eqref{A.19} have the same functional form when rewritten in terms of $\widetilde{\nabla}_a$ and $ \widetilde{\vf}$, 
\bea
( \d_{\cK[\tilde\x]} + \d_{\s[\tilde \x]} )\widetilde{\nabla}_a= 0~, \qquad
\x \widetilde{\vf} + \hf (d-2)\s[\tilde{\x }] \widetilde{\vf} =0~.
\label{2.21}
\eea

The Weyl invariance allows us to
choose the gauge condition
\bea
 \vf =1~.
 \eea
Then  the Killing equations \eqref{2.21} turn into 
\bea
 \Big[ \x^b \nabla_b +\hf K^{bc} [\x] M_{bc} , \nabla_a\Big]  &=&0~, \qquad
 \s[\x] = 0 ~,
 \eea
which is equivalent to the standard Killing equation
\bea
 \nabla^a \x^b + \nabla^b \x^a = 0~.
 \eea
 

\section{(Conformal) symmetries of curved superspace}

The Weyl invariant approach to gravity and spacetime symmetries, 
which was reviewed in the previous section, has a natural superspace
extension \cite{BK,K13,KLRST-M,KNT-M}
in all cases when Poincar\'e or anti-de Sitter supergravity is formulated as 
conformal supergravity coupled to certain conformal compensator(s).
This is always possible for supergravity theories in $d\leq 6$ 
with up to eight supercharges, where off-shell conformal supergravity always exists.  

Let $\cM^{d|\d}$ be a curved superspace, with $d$ spacetime 
and $\d$ fermionic dimensions, chosen to describe a given supergravity theory.
We denote by  $z^M = (x^m, \q^{\hat \m}) $ the local coordinates
for  $\cM^{d|\d}$. Without loss of generality, we assume that 
the zero  section of $\cM^{d|\d}$  defined by $\q^{\hat \m} =0$ corresponds to 
the spacetime manifold $\cM^d$. 

The  differential geometry of curved superspace $\cM^{d|\d}$ may be realised 
in terms of covariant derivatives  of the form 
\bea
\cD_A = (\cD_a, \cD_{\hat \a}) = E_A +\O_A +\F_A ~.
\label{3.1}
\eea
Here $E_A = E_A{}^M (z)\pa / \pa z^M $ denotes the inverse superspace vielbein,
$\O_A = \hf \O_A{}^{bc} (z) M_{bc} $ is the  Lorentz connection, 
and $\F = \F_A{}^I(z) J_I $ the $R$-symmetry connection.\footnote{The superspace 
structure group,  ${\rm Spin}(d-1,1) \times G_R$, is 
a subgroup of the isometry group of Minkowski superspace ${\mathbb R}^{d|\d}$. 
This subgroup is the isotropy group of the origin in ${\mathbb R}^{d|\d}$.}
The index $\hat \a$ of the fermionic operator $\cD_{\hat \a}$ is, in general,  composite; 
it is comprised of a spinor index $\a$  and an $R$-symmetry index. 

The covariant derivatives obey the (anti-)commutation relations of the form
\bea
[ \cD_A , \cD_B\} = \cT_{AB}{}^C \cD_C + \hf \cR_{AB}{}^{cd} M_{cd} 
+\cR_{AB}{}^I J_I~, 
\label{3.2}
\eea
where $ \cT_{AB}{}^C (z)$ is the torsion tensor, $ \cR_{AB}{}^{cd} (z)$ and 
$\cR_{AB}{}^I (z)$ are the Lorentz and $R$-symmetry curvature tensors, respectively. 
In order to describe conformal supergravity, the superspace torsion $\cT_{AB}{}^C$
has to obey certain algebraic constraints, which may be thought of 
as generalisations of the torsion-free constraint in gravity, 
eq. \eqref{torsion-free}, and which have to be Lorentz and $R$-symmetry invariant. 
Unlike the gravity case, there is no universal expression for such constraints, 
since their explicit form depends on the dimension of spacetime $d$ 
as well as on the supersymmetry type chosen. However, 
certain guiding principles leading to proper torsion constraints 
are described in important papers by Gates et al. \cite{GSW} and Howe \cite{Howe},
 and are also reviewed in textbooks \cite{BK,GGRS}.

The supergravity gauge group includes a subgroup  generated by local transformations
\begin{subequations}
\bea
\d_\cK \cD_A &=& [\cK , \cD_A] ~, \qquad 
\qquad \cK:= \x^B (z) \cD_B + \hf K^{bc} (z) M_{bc} + K^I (z) J_I ~, 
\label{3.3a}
\eea
where the gauge parameters $\x^A$, $K^{bc}= -K^{cb}$ and $K^I$
obey standard reality conditions but are  otherwise  arbitrary. 
Given a tensor superfield $\F(z)$ (with suppressed Lorentz and $R$-symmetry 
indices), its transformation law under \eqref{3.3a} is
\bea
\d_\cK \F &=& \cK \F~.  
\eea
\end{subequations}

In order to describe conformal supergravity, the constraints imposed on the superspace 
torsion should be invariant under super-Weyl transformations of the form
\begin{subequations}
\bea
\d_\s \cD_a &=& \s \cD_a + \cdots~, \qquad \d_\s \cD_\hal = \hf \s \cD_\hal + \cdots~, 
\label{3.4a}
\eea
where the scale parameter $\s$ is an arbitrary real superfield. The ellipsis 
in the expression for $\d_\s \cD_a $ includes, in general, 
a linear combination of the spinor covariant derivatives $\cD_{\hat \b}$ and 
the structure group generators $M_{cd}$ and $J_K$.  The ellipsis in 
$\d_\s \cD_\hal $ stands for a linear combination of the generators
of  the structure group. The super-Weyl transformation \eqref{3.4a} is 
a natural generalisation of the Weyl transformation \eqref{A.5} in gravity.
In most cases of interest, matter superfields may be chosen to be primary under the super-Weyl group, 
\bea
\d_\s \F &=& w_\F \s \F~,
\eea
\end{subequations}
with $w_\F$ a  super-Weyl weight. The transformations \eqref{3.3a} and \eqref{3.4a} 
generate the gauge group of conformal supergravity. 

An important difference between the superspace covariant derivatives \eqref{3.1}
and the spacetime ones, eq. \eqref{2.33}, is that the superspace 
structure group includes not only the Lorentz group, 
but also the $R$-symmetry group $G_R$.  In principle, it is always possible 
to deal with an alternative superspace geometry such that its structure group coincides
with the Lorentz group, similar to the Wess-Zumino formulation \cite{WZ}
of 4D $\cN=1$ supergravity. The local $G_R$ group will then appear as an 
additional invariance of the superspace constraints 
(similar to the (super-)Weyl invariance in (super)gravity). In many cases, however, such a formulation is technically less useful due to the presence of dimension-1/2 constraints, 
as explained by Howe \cite{Howe} in the four-dimensional case.  
 
It should be mentioned that there exist alternative approaches to
conformal gravity and conformal supergravity.
Conformal gravity in $d$ dimensions
can be formulated as a gauge theory of the conformal group, 
see, e.g., \cite{BKNT-M1} for a review. 
In such a formulation, the local special conformal transformations 
may be used to gauge away the dilatation connection. 
This will lead to the realisation for conformal gravity described in the previous section. 
Analogously, conformal supergravity in diverse dimensions $d\leq 6$ 
can be obtained by gauging the relevant
superconformal group in superspace \cite{Butter,BKNT-M1,BKNT-M3}. 
The resulting formulation, known as conformal superspace, may be viewed as a superspace version of the superconformal tensor calculus, see, e.g., \cite{FVP} for a review. The formulation for conformal supergravity described above 
is obtained from conformal superspace by gauge fixing certain local symmetries. 
It is completely adequate to study (conformal) isometries of curved 
superspace backgrounds; this is why we will not discuss conformal superspace here.

To describe  Poincar\'e or anti-de Sitter supergravity theories, 
the conformal supergravity  multiplet has to be coupled to 
some off-shell conformal compensators that will be symbolically denoted $\X$. 
In general, the compensators are Lorentz scalars,  and at least one of them 
has to have a  non-zero super-Weyl 
weight $w_\X \neq 0$,
\bea
\d_\s\X=w_\X\s\X~.
\label{GenCom}
\eea
They may also transform in some  
representations of the $R$-symmetry group.
The compensators are required to be nowhere vanishing in the sense that 
the $R$-symmetry singlets $|\X|^2 $ should be strictly positive. 
Different off-shell supergravity theories correspond to
different choices of $\X$. 
The notion of conformally related superspaces can be defined in complete 
analogy with the non-supersymmetric case considered in the previous section.

Let us now fix  a background superspace. 
A real vector field $\x= \x^B E_B$  on $(\cM^{d |\d}, \cD)$ is called 
conformal Killing if 
\bea
 (\d_\cK + \d_\s) \cD_A = 0~,
 \label{3.5}
\eea
for some Lorentz $K^{bc}$, $R$-symmetry $K^I$ and super-Weyl 
$\s$ parameters. For any dimension $d\leq 6$ and any conformal 
supergravity with up to eight supercharges, the following properties hold:

\begin{itemize}

\item All parameters $K^{bc} $,  $K^I$ and $\s$ are uniquely determined in terms of $\x^B$, which allows us to write $K^{bc} = K^{bc}[\x]$, $K^I = K^I[\x]$ and 
$\s = \s[\x]$;

\item The spinor component $\x^{\hat \b}$ is uniquely determined in terms of $\x^b$;

\item The vector component $\x^b$ obeys an equation that contains all information 
about the  conformal Killing vector field. 

\end{itemize}
For example, in the case of $\cN=1$ supergravity in four dimensions 
the equation on $\x^b$ reads \cite{BK}
\bea
\cD_{(\a} \x_{\b) \bd} = \bar \cD_{(\ad} \x_{\b \bd )}=0~, 
\eea
where the vector index of $\x^b$ is replaced by a pair of spinor ones, 
undotted and dotted. 
In the case of 3D $\cN=2$ supergravity studied in \cite{KLRST-M}, the equation on $\x^b$
is given by \eqref{CKSV-master} in the next section. 

By construction, the set of conformal Killing vectors on $(\cM^{d |\d}, \cD)$
is a Lie superalgebra with respect to the standard Lie bracket. This is 
the superconformal algebra of $(\cM^{d |\d}, \cD)$.
One can show that it is finite-dimensional (the argument one uses is similar 
to that described in the next section in the three-dimensional case).

Let $\x= \x^B E_B$ be a conformal Killing vector field on $(\cM^{d |\d}, \cD)$,
\begin{subequations}
\bea
(\d_{\cK [\x]} + \d_{\s [\x]}) \cD_A = 0~,  \label{3.7a}
\eea
for uniquely determined parameters
 $K^{bc}[\x]$,  $K^I[\x]$ and $\s[\x]$. 
 It is called a Killing vector field on $(\cM^{d |\d}, \cD ,\X)$ if the compensators
are invariant, 
\bea
( \d_{\cK [\x]} + w_\X \s[\x] ) \X =0~.
\label{3.7b}
\eea
\end{subequations}
 The set of Killing vectors on $(\cM^{d |\d}, \cD,\X)$
is a Lie superalgebra. 
The Killing equations \eqref{3.7a} and \eqref{3.7b} are super-Weyl invariant
in the sense that 
they hold for all conformally related superspace geometries.  

Using the compensators $\X$ we can always construct a 
superfield $\f = \f(\X)$ 
that is a singlet under the structure group and has the properties: 
(i) it is an algebraic function of $\X$; 
(ii)  it is nowhere vanishing; and 
(iii) it has a  non-zero super-Weyl weight $w_\f$, $\d_\s \f = w_\f \s \f$.
It follows from \eqref{3.7b} that  
\bea
( \d_{\cK [\x]} + w_\f \s[\x] ) \f =0~.
\label{3.9}
\eea
The super-Weyl invariance may be used to  impose the gauge condition $ \f=1$. 
Then eq. \eqref{3.9} reduces to
$ \s [\x] =0$,
and the Killing equations \eqref{3.7a} and \eqref{3.7b}  take the form 
\begin{subequations}
\bea
\big[{\cK [\x]} ,  \cD_A \big] &=& 0~,  \\
 {\cK [\x]}  \X &=&0~.
\eea
\end{subequations}

Of special interest are those  backgrounds $(\cM^{d|\d}, \cD, \X)$
which admit at least one (conformal) supersymmetry. 
Such a superspace possesses a conformal Killing vector field 
${ \x}^{{A}}$ of the type
\bea
\x^a |=0~, \qquad \x^{\hat \a} |\neq 0 ~.
\label{3.111}
\eea
Here, as always, the bar-projection 
of a superfield $U(z) = U(x, \q)$ is defined by 
$U|:= U(x, \q)|_{\q =0}$. 
We are usually interested in
purely bosonic backgrounds with the property that 
all fermionic components of the superspace torsion and curvature tensors, 
eq. \eqref{3.2},  
have vanishing bar-projections,
\bea
\ve(\cT_{\cdots}{}^{\cdots}) = {1} \rightarrow \cT_{\cdots}{}^{\cdots}|={0}~, \qquad
\ve(\cR_{\cdots}{}^{\cdots}) = 1 \rightarrow \cR_{\cdots}{}^{\cdots}|=0~,
\label{3.122}
\eea
where $\e$ denotes the Grassmann parity, 
$\e =0$ for bosons and $\e=1$ for fermions. 
If $\x^A$ is a Killing vector field with $\s[\x] =0$, 
then the bosonic requirements \eqref{3.122} naturally arise as consistency conditions. 
Indeed, let us suppose that $\cB$ is a bosonic part, $\ve(\cB) =0$,
 of the superspace torsion or curvature.
For $\s[\x] =0$, the transformation of  $\cB |$ is 
$ \d \cB | = \cK[\x]  \cB | = \x^{\hat \a} |\cD_{\hat \a} \cB |$, assuming all other bosonic 
parameters, $K^{bc}[\x] |$ and  $ K^I[\x]|$, vanish.  
On the other hand, it must hold that 
$\d \cB|=0$, since the geometry does not change under the transformation 
associated with $\x^A$. 
This is consistent provided $\cD_{\hat \a} \cB | =0$, which indicates that
all fermionic components of the superspace torsion and curvature tensors should vanish.

The conditions \eqref{3.122} imply that  at the component level
all fermionic fields may be gauged away.
In particular, the background gravitini are purely gauge degrees of freedom.


\section{Backgrounds with (conformal) isometries in 3D $\cN=2$ supergravity}

As an application of the formalism described in the previous section, 
we  review the results of
\cite{KLRST-M} devoted to the construction of supersymmetric 
backgrounds in  all known off-shell formulations for 
3D $\cN=2$ supergravity \cite{KT-M11,KLT-M11}.
We consider a curved superspace  in three spacetime dimensions, 
$\cM^{3|4}$,  parametrised by
local bosonic ($x^m$) and fermionic ($\q^\m, \bar \q_\m$)
coordinates  $z^{{M}}=(x^{m},\q^{\mu},{\bar \q}_{{\mu}})$,
where $m=0,1,2$ and  $\mu=1,2$.
The Grassmann variables $\q^{\mu} $ and $\bar \q_{{\mu}}$
are related to each other by complex conjugation:
$\overline{\q^{\mu}}=\bar \q^{{\mu}}$.

\subsection{$\cN=2$ conformal supergravity in three dimensions}

As discussed in section 2, conformal gravity can be described in terms of
the frame field $e_a = e_a{}^m (x) \pa_m$ defined modulo the  gauge transformations
\eqref{A.10a}. Here we review the generalisation of that formulation to the case of 
3D $\cN=2$ conformal supergravity,  following \cite{HIPT,KLT-M11,KT-M11}. 

The superspace structure group is chosen to be ${\rm SL}(2,{\mathbb{R}})\times {\rm U(1)}_R$,
and the covariant derivatives
$\cD_{{A}} =(\cD_{{a}}, \cD_{{\a}},\bar \cD^\a)$
have the form
\bea
\cD_{{A}}&=&E_{{A}}
+\O_{{A}}
+\ri \,\F_{{A}} J~,
\label{CovDev}
\eea
with $J$ the $R$-symmetry generator.
The Lorentz connection can be written in three different forms,
\bea
\O_A=\hf\O_{A}{}^{bc}M_{bc}
=\hf\O_{A}{}^{\b\g}M_{\b\g}
=-\O_A{}^c M_c~,
\label{Lorentzconnection}
\eea
depending on whether we use
the Lorentz generators with two  vector indices ($M_{ab}=-M_{ba}$),
one vector index ($M_a$) and two  spinor indices
($M_{\a\b}=M_{\b\a}$).\footnote{These generators
are related to each other as follows:
$M_{a}=\hf\ve_{abc}M^{bc}$,
$M_{ab}=-\ve_{abc}M^c$,
$M_{\a\b}=(\g^a)_{\a\b}M_{a}$ and 
$M_{a}=-\hf(\g_a)^{\a\b}M_{\a\b}$.}
The $R$-symmetry 
and  Lorentz generators act on the covariant derivatives  as follows:
\begin{subequations}
\bea
&{[} J,\cD_{\a}{]}
=\cD_{\a}~,
\qquad
{[} J,\cDB^{\a}{]}
=-\cDB^\a~,
\qquad
{[}J,\cD_a{]}=0~, \\
&{[}M_{\a\b},\cD_{\g}{]}
=\ve_{\g(\a}\cD_{\b)}~,\qquad
{[}M_{\a\b},\cDB_{\g}{]}=\ve_{\g(\a}\cDB_{\b)}~,
~~~
{[}M_{ab},\cD_c{]}=2\eta_{c[a}\cD_{b]}~.
\label{generators}
\eea
\end{subequations}
The supergravity gauge group 
includes
local $\cK$-transformations
of the form
\be
\d_\cK \cD_{{A}} = [\cK  , \cD_{{A}}]~,
\qquad \cK = \x^{{C}} \cD_{{C}} +\hf K^{ c d } M_{c d}
+\ri \, \t  J  ~,
\label{tau}
\ee
with the gauge parameters
obeying natural reality conditions, but otherwise  arbitrary.

In order to describe $\cN=2$ conformal supergravity, the torsion 
has to obey the covariant constraints proposed in \cite{HIPT}.
The resulting algebra of covariant derivatives is \cite{KLT-M11,KT-M11}
\begin{subequations} \label{algebra-final}
\bea
\{\cD_\a,\cD_\b\}
&=&
-4\bar{\cR}M_{\a\b}
~,~~~~~~
\{ \bar \cD_\a, \bar \cD_\b \}
= 4\cR M_{\a\b}~,
\\
\{\cD_\a, \bar \cD_\b \}
&=&
-2 \ri (\g^c)_{\a\b} \cD_c
-2\cC_{\a\b} J
-4\ri \ve_{\a\b}\cS J
+4\ri\cS M_{\a\b}
-2\ve_{\a\b}\cC^{\g\d} M_{\g\d}
~.
\label{4.3b}  
\eea
\end{subequations}
The explicit expressions for commutators 
$[ \cD_{a},\cD_\b ]$, ${[}\cD_{a}, \bar \cD_\b{]}$ and ${[}\cD_{a},\cD_b{]}$ are given 
in \cite{KT-M11} and \cite{KLRST-M}. 
The algebra involves three  dimension-1 torsion superfields:
a real scalar $\cS$, a
complex scalar $\cR$ and its conjugate $\bar{\cR}$,  and a real vector $\cC_a$.
The ${\rm U(1)}_R$ charge of $\cR$ is $-2$. 
The torsion superfields obey certain constraints implied by the Bianchi identities. 
Some of these constraints are
\begin{subequations}
\bea
\cDB_\a \cR&=&0~,
\\
(\cDB^2-4\cR)\cS 
&=&0~, \qquad \bar \cS = \cS~.  \label{2.12b}
\eea
\end{subequations}
Thus $R$ is covariantly chiral, and $\cS$ covariantly  linear.

The algebra of covariant derivatives given by \eqref{algebra-final}
does not change under the super-Weyl transformation
\cite{KT-M11,KLT-M11}
\bsubeq  \label{2.3}
\bea
\cD'{}_\a&=&\re^{\hf\s}\Big(\cD_\a+(\cD^{\g}\s)
M_{\g\a}-(\cD_{\a }\s)J\Big)~,
\\
\cD'{}_{a}
&=&\re^{\s}\Big(
\cD_{a}
-\frac{\ri}{2}(\g_a)^{\g\d}(\cD_{\g}\s)\cDB_{\d}
-\frac{\ri}{2}(\g_a)^{\g\d}(\cDB_{\g}\s)\cD_{\d} \non \\
&&~~~~~
+\ve_{abc}(\cD^b\s)M^c
-\frac{\ri}{2}(\cD^{\g}\s)(\cDB_{\g}\s)M_{a}
\non\\
&&~~~~~
-\frac{\ri}{8}(\g_a)^{\g\d}({[}\cD_{\g},\cDB_{\d}{]}\s)J
-\frac{3\ri}{4}(\g_a)^{\g\d}(\cD_{\g}\s)(\cDB_{\d}\s)J
\Big)~,
\eea
which induces the following transformation of the torsion tensors:
\bea
\cS'&=&\re^{\s}\Big(
\cS
+\frac{\ri}{4}\cD^\g\cDB_{\g}\s
\Big)~,
 \label{2.11d} \\
\cC'_{a}&=&\re^{\s}\Big(
\cC_{a}
+\frac{1}{8}(\g_a)^{\g\d}  [\cD_{\g},\cDB_{\d}]\s
+\frac{1}{4}(\g_a)^{\g\d}(\cD_{\g}\s)\cDB_{\d}\s
\Big)~,
\\
\cR' &=&
\re^{\s}\Big(
\cR
+\frac{1}{4} \cDB^2\s
-\frac{1}{4}  ( \cDB_\g\s)\cDB^{\g}\s
\Big)~. \label{2.11f}
\eea
\esubeq
Here the  parameter $\s$ is an arbitrary real scalar superfield. 
The infinitesimal version of super-Weyl transformation \eqref{2.3}
provides a concrete realisation of \eqref{3.4a}.

The gauge group of  conformal supergravity 
is defined to be spanned by the $\cK$-transformations \eqref{tau} 
and the super-Weyl transformations. 
The super-Weyl invariance is the reason why the  superspace 
geometry introduced describes the conformal supergravity  multiplet. 

Using the super-Weyl transformation laws \eqref{2.3}, 
one may check that the real symmetric spinor superfield \cite{Kuzenko12}
\bea
\cW_{\a\b} := \frac{\ri}{2} \big[\cD^\g ,\bar \cD_\g \big]
\cC_{\a\b} - \big[ \cD_{(\a} , \bar \cD_{\b)} \big] \cS 
- 4 \cS \cC_{\a\b}
\label{Cotton}
\eea
transforms homogeneously, 
\bea
\cW_{\a\b}' =  \re^{2\s}\, \cW_{\a\b}~.
\label{Cotton-Weyl}
\eea
This superfield is the $\cN=2$ supersymmetric generalisation of the Cotton tensor
 \bea
 W_{ab}:=\hf\ve_{acd} W^{cd}{}_{b} =  W_{ba}~, \qquad
 W_{abc}
=2{\nabla}_{[a} R_{b]c}
+\frac{1}{2}\eta_{c[a}{\nabla}_{b]}  R
\label{5.177}
\eea 
in 3D pseudo-Riemannian geometry.
A curved superspace background  $(\cM^{3|4}, \cD)$ 
is conformally flat iff the super-Cotton tensor $\cW_{\a\b}$ 
vanishes \cite{BKNT-M1}.  

\subsection{Compensators} \label{subsection4.2}

In order to describe 3D $\cN=2$ Poincar\'e or anti-de Sitter supergravity theories, 
the conformal supergravity multiplet 
has to be coupled to a certain conformal compensator $\X$ and its conjugate. 
In general, $\X$ is a scalar superfield of super-Weyl weight $w\neq 0$ and U($1)_R$ charge $q$, 
\bea
\d_\s\X=w\s\X~,\qquad 
J\X=q\X ~,
\eea
chosen to be nowhere vanishing, $\X \neq 0$. It  is assumed that $q=0$ if and only if $\X$
is real, which is the case for $\cN=2$ supergravity with a real linear compensator 
(see below). 
Different off-shell supergravity theories correspond to
different superfield types of $\X$. 
 
Type I minimal supergravity \cite{KT-M11,KLT-M11}
is a 3D analogue of the old minimal formulation 
for 4D $\cN=1$ supergravity \cite{WZ,old}. It 
makes use of two compensators, a
covariantly chiral scalar $\J$ and its conjugate $\bar \J$ with the properties
\bea
\bar \cD_\a \J =0~, \qquad \d_\s \J = \hf \s \J~,\qquad J \J = -\hf \J~.
\eea
The freedom to perform the super-Weyl and local 
U(1)$_R$ transformations allows us to choose a gauge
$\J=1$,
which implies the consistency  conditions
\bea
 \cS=0~, \qquad
\F_\a=0~,\qquad
\F_{a}= \cC_{a}~.
\label{4.133}
\eea
This reduces the superspace structure group from
${\rm SL}(2,{\mathbb{R}})\times {\rm U(1)}_R$ to its subgroup 
${\rm SL}(2,{\mathbb{R}})$.

Type II minimal supergravity \cite{KT-M11,KLT-M11}
is a 3D analogue of the new minimal formulation 
for 4D $\cN=1$ supergravity \cite{new}.
It makes use of a real 
covariantly  linear compensator $\mathbb G$ with the properties
\bea
(\bar \cD^2 -4\cR)\mathbb G = (\cD^2 -4\bar \cR)\mathbb G  =0~, 
\qquad \d_\s \mathbb G = \s \mathbb G ~.
\eea
The super-Weyl invariance allows us to choose the gauge
$\mathbb{G}=1$, which implies 
\bea
\cR=\bar{\cR}=0 ~.
\label{4.155}
\eea
Unlike the 4D case, this formulation is suitable to describe anti-de Sitter supergravity
\cite{KT-M11}.

Non-minimal $\cN=2$ Poincar\'e supergravity \cite{KT-M11,KLT-M11}
is a 3D analogue of the  non-minimal 4D $\cN=1$ supergravity 
(see \cite{BK,GGRS} for reviews).
It makes use of
a complex covariantly linear superfield $\S$ and its conjugate $\bar \S$. 
The superfield  $\S$ is characterised by the properties  \cite{KLT-M11} 
\bea 
(\cDB^2-4\cR)\S=0~, \qquad \d_\s \S=w\s\S ~,\qquad J\S=(1-w) \S~,
\label{complex-linear}
\eea
for some real parameter $w$. No reality condition is imposed on $\S$. 
The only way to describe anti-de Sitter supergravity using a non-minimal
formulation \cite{KT-M11} (in complete analogy with the four-dimensional $\cN=1$ 
case \cite{BKdual})
consists in choosing $w=-1$ in \eqref{complex-linear}
and replacing the  constraint $(\cDB^2-4\cR)\S=0$ with a deformed one, 
\begin{align}
-\frac{1}{4} (\bar \cD^2 - 4 \cR) \Gamma = \m ={\rm const}~.
\end{align}
The freedom to perform the super-Weyl and local 
U(1)$_R$ transformations allows us to choose a gauge
$\S=1$.  

Supersymmetric spacetimes in non-minimal supergravity are analogous 
to (but more restrictive than) those in Type I supergravity \cite{KLRST-M}. This is why 
we will not consider non-minimal supergravity in what follows.

\subsection{Conformal Killing vector fields on $(\cM^{3 |4}, \cD)$}

Let $\x = \x^A E_A$ be a  real vector field on
 $(\cM^{3|4}, \cD)$, with $\x^A  \equiv (\x^a , \x^\a , \bar \x_\a) $.
 It is conformal Killing  provided eq. \eqref{3.5} holds. 
  Since the vector covariant derivative $\cD_a$ is given in terms of an anti-commutator of two spinor ones, eq. \eqref{4.3b},   it suffices to analyse the  implications of 
\bea
(\d_\cK + \d_\s) \cD_\a =0~.
\label{4.5}
\eea
We should stress that 
the other requirement contained in \eqref{3.5}, 
\bea
(\d_\cK + \d_\s) \cD_{a} =0~,
\label{4.55}
\eea
is automatically satisfied provided \eqref{4.5}  holds. 

The left-hand side of \eqref{4.5} is a linear combination of the five 
 linearly independent
operators $\cD^\b$, $\bar \cD^\b$, $\cD^{\b\g}$, $M^{\b \g}$ and $J$. 
Therefore, eq. \eqref{4.5} gives five different equations. 
Let us consider in some detail the equations associated 
with the operators $\cD^\b$ and $\cD^{\b\g}$, which are
\bsubeq \label{8.3}
\bea
\cD_{{\a}} \x_{\b}
&=&
- \hf\ve_{\a\b}\big(\s
+2\ri \t\big)
 -\ri\x_{(\a}{}^{\g} \cC_{\b)\g}
+\x_{\a\b} \cS
+\hf K_{\a\b}
~, \label{8.3a}
\\
\cD_\a\x_{\b\g}&=&4\ri\ve_{\a(\b}\bar{\x}_{\g)}
~, \label{8.3b}
\eea
\esubeq
and their complex conjugate equations.
These relations imply that the  parameters $\x^\a,\,\bar{\x}_\a,\, K_{\a\b},\,\s$ and $\t$ 
are uniquely expressed in terms of  
$\x^a$ and its covariant derivatives as follows:
\bsubeq  \label{master-def}
\bea
\x^{\a}&=& - \frac{\ri}{6}\cDB_\b\x^{\b \a}
~,~~~~~~
\bar{\x}_{\a}=
-\frac{\ri}{6}\cD^\b\x_{\b\a}
 ~,
 \label{master-def-x}
\\
\s [\x]&=&
\hf\big(
\cD_{\a} \x^{\a}
+\cDB^{\a} \bar{\x}_{\a}
\big) 
~,~~~
 \label{master-def-s}
 \\
\t  [\x]
&=&
-\frac{\ri}{4}\big(
\cD_{\a} \x^{\a}
-\cDB^{\a} \bar{\x}_{\a}
\big) 
~,
 \label{master-def-t}
\\
K_{\a\b} [\x]
&=&
 \cD_{(\a} \x_{\b)}
 -\cDB_{(\a} \bar{\x}_{\b)}
-2\x_{\a\b} \cS
~.
 \label{master-def-K}
\eea
\esubeq
In accordance with \eqref{8.3b}, the remaining vector parameter $\x^a$ satisfies the 
equation\footnote{The equation \eqref{CKSV-master} is analogous 
to the conformal Killing equation, $\nabla_{(\a \b} V_{\g \d )} =0$,
on a pseudo-Riemannian three-dimensional manifold.} 
\bea
\cD_{(\a}\x_{\b\g)} &=&0 
\label{CKSV-master}
\eea
and its complex conjugate.
${}$From \eqref{CKSV-master}
one may deduce the conformal Killing equation
\bea
\cD_{a}\x_{b} + \cD_{b}\x_{a} = \frac{2}{3} \eta_{ab} \cD^{c}\x_{c}~.
\eea

Eq. \eqref{CKSV-master} is fundamental in the sense that it  implies 
$(\d_\cK + \d_\s) \cD_A \equiv 0$ provided 
the parameters $\x^\a$, $K_{\a\b}$, $\s$ and $\t$ are defined as in 
\eqref{master-def}. 
Therefore, every conformal Killing vector field  on $(\cM^{3 |4}, \cD)$ is a real vector field
\bea
\x = \x^A E_A ~, \qquad \x^A  = (\x^a , \x^\a , \bar \x_\a) :=
 \Big( \x^a , - \frac{\ri}{6}\cDB_\b\x^{\b \a} , 
-\frac{\ri}{6}\cD^\b\x_{\b\a} \Big) ~,
\eea
which obeys the master equation \eqref{CKSV-master}. 
If $\x_1$ and $\x_2$ are two conformal Killing vector fields, their 
Lie bracket $[\x_1, \x_2]$ is a conformal Killing vector field. 

The equation  \eqref{4.5} implies some additional results
that have not been discussed above.
Defining
$ \U:= \big\{ \x^B, K^{\b\g} [\x] ,  \t [\x] , \s [\x], \cD_B \s [\x] \big\} $, 
it turns out that the descendants  $\cD_A \U $
are linear combinations of the elements of $\U$.
This means that the Lie superalgebra of conformal Killing vector fields 
on $(\cM^{3|4}, \cD)$ is finite dimensional. 
The number of its even and odd generators cannot exceed those in 
the $\cN=2$ superconformal algebra  ${\mathfrak{osp}}(2|4)$.

\subsection{Killing vector fields on $(\cM^{3 |4}, \cD, \X)$}

A conformal Killing vector field $\x = \x^A E_A$  on 
$(\cM^{3|4}, \cD)$  is said to be a Killing vector field on $(\cM^{3|4}, \cD, \X)$
if the following conditions hold:
\begin{subequations} \label{8.13both}
\bea
\Big[\x^B \cD_B + \hf K^{bc}[\x] M_{bc} + \ri \t [\x] J, \cD_A \Big]  + \d_{\s [\x]} \cD_A &=&0~ ,
\label{8.13a} \\ 
\Big(\x^B \cD_B  + \ri q \t  [\x]  + w \s [\x] \Big) \X& =&0~, \label{8.13b}
\eea
\end{subequations}
with the parameters $ K^{bc}[\x]$, $\t [\x]$ and $\s[\x]$ defined as in \eqref{master-def}.
The set of all Killing vector fields on  $(\cM^{3|4}, \cD, \X)$  is a Lie superalgebra. 
The Killing vector fields generate the symmetries 
of rigid  supersymmetric  field theories defined on this superspace. 

The Killing equations \eqref{8.13both} are super-Weyl invariant 
in the sense that they have the same form for conformally related superspaces. 
The super-Weyl and local U(1)$_R$ symmetries allow us
to choose the useful gauge
\bea
\X=1~,
\label{Xgauge1}
\eea
which characterises the off-shell supergravity formulation chosen.
If $q \neq 0$, there remain no residual super-Weyl and local U(1)$_R$ 
 symmetries in this gauge. 
If $q=0$, the local U(1)$_R$ symmetry remains unbroken while the super-Weyl freedom 
is completely fixed.

In the gauge \eqref{Xgauge1}, the Killing equation \eqref{8.13b} becomes
\bea
\ri q\Big(\x^B\F_B
+\t [\x] \Big)+w\s[\x]  = 0 ~,
\eea
where $\F_B$ is the U$(1)_R$ connection, eq. \eqref{CovDev}.
Hence, the isometry transformations are generated by those conformal Killing supervector
fields which respect the conditions
\bsubeq
\bea
\s [\x]&=&0
~,
\label{constr-sigma}
\\
q\neq 0 \quad \Longrightarrow \quad \t [\x]&=&
- \x^B \F_B ~.
\label{constr-tau}
\eea
\esubeq
These properties provide the main rationale for choosing the gauge condition  \eqref{Xgauge1}
which is:
for any off-shell supergravity formulation, the isometry transformations 
are characterised by the condition $\s [\x]=0$, which eliminates 
super-Weyl transformations.

\section{Supersymmetric three-dimensional spacetimes}

Let us look for curved superspace backgrounds $(\cM^{3 |4}, \cD)$ which  admit
at least one conformal supersymmetry. 
Such a superspace must possess 
a conformal Killing vector field  with the property
\bea
\x^a|=0 ~, \qquad \e^\a (x) := \x^\a | \neq 0~.
\eea
All other bosonic parameters are assumed to vanish, 
$\s | =  \t | = K_{\a\b}|=0$. Then any parameter of the type 
$(\cD_{B_1} \cdots \cD_{B_n} \x^A)|$ is expressed in terms of 
the two spinor parameters:
$Q$-supersymmetry $\e^\a(x)$ and $S$-supersymmetry
 $ \eta_\a (x) := \cD_\a \s|$.
This follows from the general properties of the conformal Killing vector fields 
on $(\cM^{3|4}, \cD)$ discussed above. 

In the 3D $\cN=2$ case, all bosonic superspace  backgrounds,
which possess no covariant fermionic fields, 
are characterised by the conditions:
\bea
\cD_\a\cS|= 0~, \qquad \cD_\a \cR|= 0~, \qquad \cD_\a \cC_{\b\g}|=0~.
\label{5.22}
\eea
These conditions mean that the gravitini can be gauged away
such that 
\bea
\cD_a|=\bD_a :=
e_a{}^m (x) \pa_m
+\hf\o_a{}^{bc} (x) M_{bc}
+\ri b_a (x) J = {\nabla}_a +\ri b_a (x) J ~,
\label{5.33}
\eea
where $\nabla_a$ stands for  the torsion-free covariant derivative \eqref{2.33}. 
 Introduce scalar and vector fields associated with the superspace torsion:
\bea
{s} (x) := \cS|~, \qquad {{r}} (x) := \cR|~, 
\qquad {c}_a (x) := \cC_a|~.
\eea

The spinor parameter $\e = (\e_\a) $ proves to obey the equation
\bea
\bD_{a}\e
+\frac{\ri}{2}{\g}_a\bar{\eta}
+\ri\ve_{abc}\,{c}^b {\g}^c\e-{s} {\g}_a\e-\ri{{r}}{\g}_a\bar{\e} =0
~.
\label{4.14} 
\eea
This equation is obtained by bar-projecting the relation
\bea
0=
\cD_{a}\x_{\a}
&+&\frac{\ri}{2}(\g_a)_{\a}{}^{\b} \cDB_{\b}\s
-\ri\ve_{abc}(\g^b)_\a{}^{\b}\cC^c\x_{\b}
-(\g_a)_\a{}^{\b}( \x_{\b}\cS
+\bar{\x}_{\b}R)
\non\\
&
+&\hf\ve_{abc}\x^{b}(\g^c)^{\b\g}\Big(
 \bar \cD_{(\a}\cC_{\b\g )}
+\frac{4\ri}{3}\ve_{\a(\b}  \cDB_{\g)}\cS 
+\frac{2}{3}\ve_{\a(\b} \cD_{\g)} R
\Big)
~,
\eea
which is one of the implications of \eqref{4.55}. 
We recall that \eqref{4.55} is automatically satisfied if the equation 
\eqref{4.5} holds. 

Eq. \eqref{4.14} 
contains two pieces of information. Firstly, it allows one to express 
the spinor parameter $\bar \eta = (\bar \eta_\a ) $ via $ \e$, 
its conjugate $\bar \e$ and covariant derivative $\bD_a\e$: 
\begin{subequations}
\bea
\bar{\eta}_\a&=&
-\frac{2\ri}{3}
\Big(
(\g^a\bD_a\e)_\a
+2\ri(\g^a\e)_\a c_a
+3{s} \e_{\a}
+3\ri {{r}}\bar{\e}_{\a}
\Big)
~.
\eea
Secondly, it gives a closed-form equation on $\e$:
\bea
&&
\Big(\bD_{(\a\b}
-\ri{c}_{(\a\b} \Big)
\e_{\g)} = 
\Big( 
\nabla_{(\a\b}
 -\ri (b + c)_{ (\a\b } \Big)
\e_{\g)}
=0
~.
\label{conf-kill-spinor}
\eea
\end{subequations}
Equation \eqref{conf-kill-spinor} tells us that $\e$ is 
a {\it charged conformal Killing spinor}, 
since  \eqref{conf-kill-spinor} can be rewritten in the form \cite{HTZ}
\bea
\widetilde{\nabla}_{(\a\b} \e_{\g)} =0~, \qquad 
\widetilde{\nabla}_{a} \e := ({\nabla}_{a}  -\ri A_a)\e ~,
\label{8.23}
\eea
where $A_a =  b_{a} + {c}_{a} $.
Switching off the U(1) connection $A $ in \eqref{8.23} gives the equation for 
conformal Killing spinors. We point out that the more conventional form of writing  
\eqref{8.23} is 
\bea
(\widetilde{\nabla}_{a}   - \frac{1}{3} \g_a \g^b \widetilde{\nabla}_{b} ) \e =0~.
\eea

Choose $\e_\a$ to be  a bosonic (commuting) spinor. 
Then, by analogy with, e.g.,  the 5D analysis in \cite{GG1},  
we deduce from \eqref{8.23} that the real vector field $V_{a} :=  (\g_a)^{\a\b}\bar \e_{\a} \e_{\b} $ has the following properties: (i) $V_a$ is a conformal Killing vector field, 
${\nabla}_{(\a\b} V_{\g\d)} =0$; and (ii) $V_a$ is null or time-like, since 
$V^a V_a = (\bar \e^\a \e_\a)^2 \leq 0$. This vector field is null if and only if 
$\bar \e_\a \propto \e_\a$. These properties  
were first observed in \cite{HTZ}.

\subsection{Supersymmetric backgrounds}

As discussed in section 3, using the compensators $\X$  one can construct 
a nowhere vanishing real scalar $\f$ with the super-Weyl transformation 
 $\d_\s \f = w_\f \s \f$,  where the super-Weyl weight $w_\f$ is non-zero. 
The super-Weyl gauge freedom can be fixed by choosing the gauge $\f=1$ 
in which $\s [\x] =0$. One may choose $\f$ to be 
(i) $ \bar \J \J $ in Type I supergravity; (ii) 
$ \mathbb G $ in Type II supergravity;  
and (iii) $\bar \S \S$ in non-minimal supergravity.   
 
In the super-Weyl gauge $\f=1$, every rigid supersymmetry transformation
is characterised by 
\bea
\s[\x] =0 \quad \Longrightarrow \quad  \eta_\a =0
~. 
\eea
Then the conformal Killing spinor equation
 \eqref{4.14} turns into 
\bea
\bD_{a} \e =
-\ri\ve_{abc}  {c}^b {\g}^c\e
+ {s}{\g}_a\e
+\ri r  {\g}_a\bar{\e} ~.
\label{4.20}
\eea
We recall that the covariant derivative 
$\bD_a$ is defined by \eqref{5.33}. 
It contains a ${\rm U(1)}_R$ connection,
and the algebra of covariant derivatives is
\bea
[\bD_a,\bD_b]= \hf  R_{ab}{}^{cd} M_{cd} +\ri  F_{ab} J 
= [{\nabla}_a , {\nabla}_b] +\ri  F_{ab} J 
~.
\eea
Eq. \eqref{4.20} is a generalised Killing spinor equation. Along with 
the frame field $e_a = e_a{}^m(x) \pa_m$, it involves four other background
fields, which are: the U$(1)_R$ gauge connection $b_a (x)$, the vector field 
$c_a (x)$, the real scalar field $s(x)$ and the complex scalar one $r(x)$. 

\subsection{Maximally supersymmetric backgrounds}

The existence of rigid supersymmetries, 
i.e. solutions of the equation \eqref{4.20}, imposes non-trivial restrictions on 
the background fields. In the case of four supercharges, these restrictions
have been analysed in \cite{KLRST-M}. They are: 
\begin{subequations} \label{5.133}
\bea
{\nabla}_a 
s
&=& 0~, \qquad \bD_a r= 
({\nabla}_a  -2\ri b_a ){{r}}=0~,
\qquad
{\nabla}_{a} {c}_b=
2\ve_{abc}{c}^c {s}
~,  \label{5.133a}
\\
r\, {s}
&=&0~,
\qquad
r \,{c}_a
= 0 ~.  \label{5.133b}
\eea
\end{subequations}
It follows that $c_a$ is a Killing vector field, 
\bea
{\nabla}_{a} {c}_b + {\nabla}_{b} {c}_a =0~, 
\eea
such that $c^2 := \eta_{ab}c^a c^b = {\rm const}$.
The U$(1)_R$ field strength proves to vanish, 
\bea
F_{ab}=0~.
\eea
For the Ricci tensor
we obtain
\bea
R_{ab}
=
4\Big{[}
{c}_a {c}_b
-\eta_{ab}\big\{ c^2 +2 \big({s}^2
+\bar {{r}} {{r}}\big) \big\}
\Big{]}~. \label{5.166}
\eea
Using this result, for the Cotton tensor defined by  \eqref{5.177}
we read off the following expression:
\bea
 W_{ab}= -24s  \big{[} {c}_{a} {c}_b -\frac{1}{3}\eta_{ab} {c}^2 \big{]}~.
\eea
It is clear that the spacetime is conformally flat if $s c_a =0$.

The above restrictions are given in terms of  component fields. 
They may be recast in the language of superspace and superfields
using a 3D analogue of the 5D observation in \cite{KNT-M}. 
For any 3D $\cN=2$ supergravity background admitting four supercharges, 
if there exists a  tensor superfield $T$ such that 
its bar-projection vanishes, $T|=0$, and this condition is supersymmetric,
then the entire superfield is zero, $T=0$.
In particular, the supersymmetric conditions \eqref{5.22}
imply
\bea
\cD_\a\cS= 0~, \qquad \cD_\a \cR= 0~, \qquad \cD_\a \cC_{\b\g}=0~.
\eea
Further superfield conditions follow from \eqref{5.133}.
As follows from \eqref{Cotton}, the super-Cotton tensor takes the form
\bea
\cW_{\a\b} = 
- 4 \cS \cC_{\a\b}~.
\label{5.199}
\eea

Up to this point, no specific compensator has been chosen, 
and all the results so far obtained are applicable to 
every off-shell formulation for 3D $\cN=2$ supergravity. 
We now turn to making a specific choice of compensators.

\subsection{Maximally supersymmetric backgrounds in Type I supergravity}

As discussed in subsection \ref{subsection4.2}, 
 in Type I supergravity the super-Weyl and local 
U(1)$_R$ transformations can be used to impose the gauge
$\J=1$, which leads to the consistency conditions \eqref{4.133}. 
The corresponding Killing spinor equation is obtained from  \eqref{4.20} 
by setting $s=0$ and $b_a = c_a$, which gives
\bea
{\nabla}_a\e
&=&
\ri {c}_a\e
-\ri\ve_{abc} {c}^b  {\g}^c\e  
+\ri{{r}}{\g}_a\bar{\e} 
~.
\label{KillingSpinor-I_0}
\eea

In the case of maximally supersymmetric backgrounds, 
the dimension-1 torsion superfields obey the constraints:
\bea
\cS=0~, \qquad \cR \,{\cC}_a
= 0 ~,
\qquad
{\cD}_A  {\cR}= 0~,
\qquad
{\cD}_{A} {\cC}_b=0~.
\label{5.211}
\eea
The complete algebra of covariant derivatives is
\begin{subequations} \label{4.37}
\bea
\{\cD_\a,\cD_\b\}
&=&
-4\bar{\cR}M_{\a\b}
~,~~~~~~
\{\cDB_\a,\cDB_\b\}
=4{\cR}M_{\a\b}~,
\\
\{\cD_\a,\cDB_\b\}
&=&
-2 \ri (\g^c)_{\a\b} \Big(  \cD_c
-\ri  \cC_{c} J \Big)
+4\ve_{\a\b}\cC^{c}M_{c}
~, \\
{[}\cD_{a},\cD_\b{]}
&=&
\ri\ve_{abc}(\g^b)_\b{}^{\g}\cC^c\cD_{\g}
-\ri(\g_a)_{\b\g}\bar{\cR}\cDB^{\g}
~,\\
{[}\cD_{a},\cDB_\b{]}
&=&
-\ri\ve_{abc}(\g^b)_\b{}^{\g}\cC^c\cDB_{\g}
-\ri(\g_a)_\b{}^{\g}{\cR}\cD_{\g}~, \\
{[}\cD_a,\cD_b]{}
&=&4  \ve_{abc}\Big(\cC^c \cC_d
+\d^c{}_d\bar{\cR}\cR 
\Big)M^d ~.
\eea
\end{subequations}
Re-defining the covariant derivatives 
$\cD_A = (\cD_a , \cD_\a , \bar \cD^\a ) \to 
\widetilde{\cD}_A = (\cD_a -\ri \cC_a J , \cD_\a , \bar \cD^\a )$
results in a supergeometry without U$(1)_R$ curvature, which means 
that the U$(1)_R$ connection can be gauged away. 
As follows from \eqref{5.199} and \eqref{5.211}, 
the super-Cotton tensor is equal to zero, and thus 
the superspace (and spacetime) geometry is conformally flat. 

There are four different maximally supersymmetric backgrounds described 
by the superalgebra \eqref{4.37}, with $\cR $ and $\cC_a$ constrained by 
\eqref{5.211}. 
The case  $\cR\neq 0 $ and $\cC_a =0$ corresponds to 
(1,1) AdS superspace \cite{KT-M11}. The other three cases are characterised 
by $\cR=0$ and correspond to different choices for a covariantly constant vector
field $c_a (x) = \cC_a|$, which are timelike, spacelike or null. 

The existence of a covariantly constant vector field 
$c^a$ means that spacetime is decomposable
in the non-null case  (see, e.g., \cite{Stephani}).
For $c^2 \neq 0$
the spacetime is the product of a two- and a one-dimensional manifold. 
We can choose a coordinate frame $x^{ m} = (x^\hm, \z) $, 
where $\hm =1,2$, 
such that the vector field $ c^a e_a$ is proportional to $\pa/ \pa \z$
and the metric reads
\bea
\rd s_3^2 = g_{\hm \hn } (x^{\hat r})  \rd x^\hm \rd x^\hn +\k (\rd \z)^2 
= \eta_{\ha \hb} e^\ha e^\hb +\k (\rd \z)^2~, \qquad 
e^\ha:=\rd x^\hm e_\hm{}^\ha (x^\hn)~,
\eea 
where $\k =-1$  when $c^a$ is timelike, and $\k =+1$ when $c^a$ is spacelike. 
The two-dimensional metric 
$\rd s_2^2 = g_{{\hm} \hn } (x^{\hat r})  \rd x^\hm \rd x^\hn $ corresponds to a 
two-dimensional submanifold $\cN^2$ of $\cM^3 $orthogonal to $ c^a e_a$. 
We denote by ${\frak R}_{\ha \hb}$ the Ricci tensor for $\cN^2$. 
Since $c^a $ is covariantly constant, $R_{ab} c^b =0$, which means $R_{a \z} =0$. 
From \eqref{5.166} we then read off 
$R_{\ha \hb}=- 4 c^2  \eta_{\ha \hb}$. This means that the submanifold $\cN^2$ is 
(i) $S^2$  if $c^a$ is timelike; and (ii) $AdS_2$ if $c^a$ is spacelike. 
Finally, in the case that $c^a$ is null, the corresponding spacetime is a special example
of pp-waves.

\subsection{Maximally supersymmetric backgrounds in Type II supergravity}

As discussed in subsection \ref{subsection4.2}, 
 in Type II supergravity the super-Weyl invariance can be used to impose the gauge
$\mathbb G=1$, which leads to the consistency conditions \eqref{4.155}. 
The corresponding Killing spinor equation is obtained from  \eqref{4.20} 
by setting $r=0$,  
\bea
\bD_{a} \e =
-\ri\ve_{abc}  {c}^b {\g}^c\e
+ {s}{\g}_a\e~.
\label{KillingSpinor-II_0}
\eea

In the case of maximally supersymmetric backgrounds, 
the dimension-1 torsion superfields obey the constraints:
\bea
\cR=0~, 
\qquad
{\cD}_A  {\cS}= 0~,
\qquad \cD_\a \cC_b =0 \quad \Longrightarrow \quad
{\cD}_{a} {\cC}_b=
2\ve_{abc}{\cC}^c {\cS} ~,
\eea
and hence $ \cC^b \cC_b ={\rm const}$.
The corresponding algebra of covariant derivatives is
\begin{subequations}\label{4.39}
\bea
\{\cD_\a,\cD_\b\}
&=&
0
~,\qquad
\{\cDB_\a,\cDB_\b\}
=
0~,
\\
\{\cD_\a,\cDB_\b\}
&=&
-2 \ri (\g^c)_{\a\b} \Big(  \cD_c - 2\cS M_c
-\ri  \cC_{c} J \Big)
+4\ve_{\a\b}\Big( \cC^{c}M_{c}- \ri \cS J\Big)
~, \\
{[}\cD_{a},\cD_\b{]}
&=&
\ri\ve_{abc}(\g^b)_\b{}^{\g}\cC^c\cD_{\g}
+ (\g_a)_\b{}^\g \cS \cD_{\g}~, \\
{[}\cD_{a},\cDB_\b{]}
&=&
-\ri\ve_{abc}(\g^b)_\b{}^{\g}\cC^c\cDB_{\g}
+(\g_a)_\b{}^{\g}\cS \bar \cD_{\g}~, \\
{[}\cD_a,\cD_b]{}
&=&4  \ve_{abc}\Big( \cC^c \cC_d
+\d^c{}_d \cS^2
\Big)M^d ~.
\eea
\end{subequations}
The solution with ${\cC}_a =0$ corresponds to  (2,0) AdS superspace
\cite{KT-M11}.
The algebras \eqref{4.37} and \eqref{4.39} coincide under the conditions $\cR = \cS =0$. 

Curved backgrounds of the type \eqref{4.39} are  solutions to 
the  equations of motion for 
topologically massive Type II supergravity with a cosmological term.
These equations are \cite{KLRST-M}
\begin{subequations}
\bea
 \ri \cD^\a \bar \cD_\a \ln \mathbb G - 4 \cS  -  2\l  \mathbb G &=&0~, \label{5.26a}\\
\frac{1}{g} \cW_{\ab}
-\frac{1}{\mathbb G} \cD_{(\a}\mathbb G \bar \cD_{\b)} \mathbb G
 +\frac{1}{4} \big[\cD_{(\a} , \bar \cD_{\b)} \big] \mathbb G  +\cC_{\a\b} \mathbb G &=&0~.
\label{5.26b}
\eea
\end{subequations}
Here $\l$ is the cosmological constant, and $g$ the coupling constant 
appearing in the conformal supergravity action 
(Newton's constant is set equal to one). 
In the super-Weyl gauge $\mathbb G =1$
these equations turn into 
\begin{subequations}
\bea
\cS +\frac{1}{2} \l &=&0~, \\
 \frac{\ri}{2} \big[\cD^\g ,\bar \cD_\g \big]\cC_{\a\b} 
+  (g+2\l) \cC_{\a\b} &=&0~, \label{5.27b}
\eea
\end{subequations}
where we have used the explicit expression for the super-Cotton tensor \eqref{Cotton}.
For a solution with a non-vanishing $\cC_{\a\b} $ constrained by 
$ \cD_\g \cC_{\a\b} =0$,  one can satisfy eq. \eqref{5.27b} 
if the coupling constants $g$ and $\l$ are related to each other as 
\bea
g +2\l =0~.
\eea

The bosonic 
solutions of topologically massive $\cN=2$ supergravity
 with a cosmological term
were classified in \cite{Chow:2009km}. Supersymmetric spacetime \eqref{4.39} 
 is of type N (for $C_a$ null), type $\rm D_s$ (for $C_a$ spacelike) or 
$\rm D_t$ (for $C_a$ timelike) in the Petrov-Segre classification, see 
\cite{Chow:2009km} for more details.

\section{Concluding comments}

In this note we reviewed the superspace formalism to determine supersymmetric 
spacetimes from off-shell supergravity in diverse dimensions. 
For a given supergravity theory, we showed that a purely bosonic background admits
rigid supersymmetry transformations provided the corresponding curved superspace
possesses a Killing vector field of the type \eqref{3.111}. Thus the superspace must 
possess nontrivial isometries that, by construction, form a finite-dimensional supergroup. 

 Within the component approaches to
supersymmetric backgrounds in off-shell supergravity theories 
\cite{FS,Jia:2011hw,Samtleben:2012gy,Klare:2012gn,DFS,KMTZ,Liu:2012bi,Dumitrescu:2012at,Kehagias:2012fh,Festuccia3D,HTZ,deMH1,DKSS2}, 
the analysis amounts to classifying all solutions of generalised Killing spinor 
equations (such as eqs. \eqref{KillingSpinor-I_0} and \eqref{KillingSpinor-II_0}
in the case of $\cN=2$ supergravity theories in three dimensions) obtained as the 
condition for the gravitino variation to vanish. Given such a solution, 
special analysis is required to understand whether  
there exists a superalgebra to which the generators of rigid supersymmetry 
transformations belong. In the superspace setting, this issue does not occur
since the rigid supersymmetry transformations belong to the isometry group
of the background superspace. 

The superspace formalism provides a simple geometric realisation 
for maximally supersymmetric spacetimes. They correspond 
to those off-shell supergravity backgrounds for which the Grassmann-odd components 
of the superspace torsion and curvature tensors vanish, while the Grassmann-even components of these tensors are annihilated by the spinor derivatives. 
This follows from the observation that, 
for every maximally supersymmetric background, 
if there exists a  tensor superfield $T$ such that 
its bar-projection vanishes, $T|=0$, and this condition is supersymmetric,
then the entire superfield is zero, $T=0$. 
As a simple corollary of this result, one can readily deduce that 
all maximally supersymmetric spacetimes 
are conformally flat for certain supergravity theories. For instance, 
in the case of 4D $\cN=1$ supergravity, the super-Weyl tensor is 
a completely symmetric spinor superfield $W_{\a\b\g}$ \cite{WZ}. 
Since it must vanish for every
maximally supersymmetric spacetime, the corresponding Weyl tensor is equal to zero.  
This vanishing of the Weyl tensor was observed in \cite{FS}, but no explanation 
of this result was given. Another example is provided by 3D $\cN=1$ 
supergravity in which the super-Cotton tensor is again a
symmetric spinor superfield $W_{\a\b\g}$ \cite{KT-M12,BKNT-M1}. 
The Cotton tensor is one of the components fields 
contained in   $W_{\a\b\g}$.  Since $W_{\a\b\g}$ must vanish 
for every maximally supersymmetric 3D spacetime, 
the corresponding Cotton tensor is equal to zero.  
Our last example is provided by 3D $\cN=3$ 
supergravity in which the super-Cotton tensor is a
 spinor superfield $W_{\a}$ \cite{BKNT-M1}.
 Since $W_{\a}$ must vanish 
for every maximally supersymmetric background of $\cN=3$ supergravity, 
the corresponding 3D spacetime is conformally flat. 

A striking feature of superspace techniques is that
they make it possible to generate arbitrary off-shell supergravity-matter couplings
(such as the off-shell locally supersymmetric sigma models in 5D $\cN=1$ \cite{KT-M5D},
4D $\cN=2$ \cite{KLRT-M} and 3D $\cN=3$ and $\cN=4$ \cite{KLT-M11} supergravity theories).   
Restricting these couplings to a given background allows one to construct 
general rigid supersymmetric theories on such a spacetime.  
\\

\noindent
{\bf Acknowledgements:}\\
The author thanks Daniel Butter, Joseph Novak, 
Gabriele Tartaglino-Mazzucchelli and Arkady Tseytlin for helpful comments 
on the manuscript. 
This conference paper is based in part on joint publications with 
Ulf Lindstr\"om, Joseph Novak, Martin Ro\v{c}ek, Ivo Sachs and 
Gabriele Tartaglino-Mazzucchelli.  The author is grateful to all of them 
for enjoyable collaboration. 
This work was supported in part by the Australian Research Council
projects DP1096372 and DP140103925.

\end{document}